\documentclass[]{aspm}

\articleinfo{}{}{}

\usepackage{verbatim}
\usepackage{amssymb}
\usepackage{amsbsy}
\usepackage{amscd}
\usepackage{amsmath}
\usepackage{amsthm}
\usepackage[mathscr]{eucal}

\numberwithin{defn}{section}
\newtheorem{thm}{Theorem}

\newtheorem{dfn}{Definition}

\newtheorem{rem}{Remark}

\newcommand{\R}{\mathbb R}

\newcommand{\SSS}{\mathbb S}
\newcommand{\HH}{\mathbb H}

\numberwithin{equation}{section}

\title[Qausi-local mass]{Quasi-local mass and isometric embedding with reference to a static spacetime}

\author[]{Mu-Tao Wang}

\address{Department of Mathematics, Columbia University, 2990 Broadway, New York, NY 10027}

\email{mtwang@math.columbia.edu}

\rcvdate{}
\rvsdate{}

\subjclass[2010]{}


\begin{document}

\begin{abstract} The mathematical theory of isometric embedding is applied to study the notion of quasilocal mass in general relativity. In particular, I shall report some recent progress of quasilocal mass with reference to a cosmological spacetime, such as the de Sitter or the Anti-de Sitter spacetime, or a blackhole spacetime, such as the Schwarzschild spacetime. This article is based on joint work with Po-Ning Chen, Ye-Kai Wang, and Shing-Tung Yau. 
\end{abstract}

\maketitle

\section{Isometric embedding into $\R^3$}
 A good starting point of the theory of surface isometric embedding is Weyl's problem \cite{Weyl}:
Given a Riemannian metric $\sigma$ of positive Gauss curvature  on the two-sphere $S^2$, does there exist a closed convex surface in $\R^3$ whose first 
fundamental form is $\sigma$? In a local coordinate system $(u, v)$ on $S^2$, \[\sigma=E (u, v) du^2+2F (u, v) dudv+G(u, v) dv^2\] with $EG-F^2>0$,  we ask if there exists an embedding 
$X:S^2 \rightarrow \R^3$ such that 
\[\begin{cases}
&\langle \frac{\partial X}{\partial u}, \frac{\partial X}{\partial u}\rangle=E\\
&\langle \frac{\partial X}{\partial u}, \frac{\partial X}{\partial v}\rangle=F\\
&\langle \frac{\partial X}{\partial v}, \frac{\partial X}{\partial v}\rangle=G.\\
\end{cases}\]

 When the Gauss curvature is positive, this is a nonlinear PDE system of ``elliptic" type and the problem was solved independently by Nirenberg \cite{Nirenberg} and Pogorelov \cite{Pogorelov}.  The uniqueness of the solution was proved by Cohn-Vossen \cite{CV} in the analytic case and by Herglotz \cite{Herglotz} in the smooth
case.  Therefore, Weyl's problem always has a unique solution, i.e. any two solutions differ by a rigid motion of $\R^3$.

\section{Quasilocal mass in general relativity}
 How is the surface isometric embedding problem related to the notion of quasilocal mass in general relativity? A fundamental difficulty in general relativity is that, unlike any other physical theory, there is no mass/energy density for gravitation. This is a direct consequence of Einstein's equivalence principle.  The familiar formula that mass is the bulk integral of mass density is simply not available. In the case of Newtonian gravity, which is the predecessor of  the theory of general relativity, there is the formula $\Delta \Phi=4\pi \rho$, where $\rho$ is the mass density. For a domain $\Omega\subset \R^3$, by the divergence theorem, the total mass $\int_{\Omega} \rho$ is the same as the flux integral \[\frac{1}{4\pi}\int_{\partial\Omega} \frac{\partial\Phi}{\partial \nu}\] on the boundary surface  $\partial\Omega$.

In 1982, Penrose \cite{Penrose1} proposed a list of major unsolved problems in classical general relativity, and the first one is:
Find a suitable definition of  quasi-local mass for  a spacelike domain $\Omega$ in a general physical spacetime, as an integral over the boundary surface $\partial \Omega=\Sigma$.

There are many different approaches to this problem (Hawking \cite{Hawking}, Penrose \cite{Penrose2}, Bartnik \cite{Bartnik2}, Brown-York \cite{BY2}, Dougan-Mason\cite{Dougan-Mason}, etc.). A very promising one is the Hamilton-Jacobi analysis of gravitational action \cite{BY2, HH} which indeed leads to another fundamental difficulty 
in general relativity: there is no canonical reference system.

Applying the Hamilton-Jacobi analysis to  the Hilbert-Einstein action $\int \mathfrak{R}$, where $\mathfrak{R}$ is the scalar curvature of spacetime, Brown and York \cite{BY2} derived a surface Hamiltonian (see also Hawking-Horowitz \cite{HH}, Kijowski \cite{Ki}) on a two-dimensional spacelike surface $\Sigma$.  The expression depends on  $t^\mu$, a timelike unit vector field restricting to $\Sigma$, and $u^\mu$, a timelike unit normal vector field along $\Sigma$. One considers a spacelike domain $\Omega$ bounded by $\Sigma$ and let $v^\mu$ be the outward spacelike unit normal of $\Sigma$ that is tangent to $\Omega$. $u^\nu$ can then be taken to be the timelike normal that is orthogonal to $v^\mu$. Along $\Sigma$, decompose \[t^\mu=N u^\mu+N^\mu,\]  where $N$ is called the lapse function and $N^\mu$ is called the shift vector.  The surface Hamiltonian (up to a factor of $\frac{1}{8\pi}$) is \[\mathcal{H}(t^\mu, u^\mu)=\int_\Sigma [N k-N^\mu P_{\mu\nu} v^\nu],\] where $k$ is the mean curvature of $\Sigma$ in $\Omega$, and $P_{\mu\nu}$ is the second fundamental form of $\Omega$ in spacetime.
 It turns out the expression depends only on the geometric data $(\sigma, H)$ where $\sigma$ is the induced metric on $\Sigma$, and $H$ is the mean curvature vector field of $\Sigma$, in addition to the gauge choice of $(t^\mu, u^\nu)$ along $\Sigma$. The mean curvature vector field $H$ is the unique normal vector field of $\Sigma$ which characterizes the area expansion or variation of $\Sigma$. 

 To obtain the {\it energy}, one needs a reference Hamiltonian to define:
\[energy=\text{physical Hamiltonian}- \text{ reference Hamiltonian}.\] The reference Hamiltonian should correspond to a surface term in a reference spacetime. The theory of surface isometric embeddings is applied in finding the reference surface.  One also needs to specify ${t}^\mu$ and  ${u}^\mu$.

Brown and York \cite{BY2} fixed $\Omega$ and took $t^\mu=u^\mu$, which corresponds to $N=1$ and $N^\mu=0$. The reference surface was taken to be an isometric embedding of $\sigma$ into $\R^3$. The Brow-York mass is defined to be \[\frac{1}{8\pi} \int_{\Sigma} (H_0-k),\] where $H_0$ is the mean curvature of the image surface of the isometric embedding of $\sigma$.  The expression does converge to the ADM mass \cite{Arnowitt-Deser-Misner}under standard asymptotically flat conditions.  Another important justification of the Brown-York mass is the positivity/rigidity theorem of Shi-Tam \cite{Shi-Tam}. Yau proposed that a spacetime version is needed to study the dynamics of Einstein's equation. Liu-Yau \cite{Liu-Yau1} took the spacetime mean curvature vector field $H$ (assumed to be spacelike) of $\Sigma$ and defined the Liu-Yau mass \cite{Liu-Yau1, Liu-Yau2} (see also Kijowski \cite{Ki}, Booth-Mann \cite{Booth-Mann}, Epp \cite{Epp}) to be 
\[\frac{1}{8\pi}\int_\Sigma (H_0-|H|),\] and proved the positivity. However, these definitions turn out to be ``too positive" as there are surfaces (O'Murchadha et al \cite{OST}) in the Minkowski spacetime $\R^{3,1}$ with strictly positive Brown-York or Liu-Yau mass.

 At this point, it appeared clear that one should look for a quasilocal mass definition that uses the Minkowski spacetime as a reference spacetime. An important criterion is that the quasilocal mass of a surface in the Minkowski spacetime should be zero, as there is no matter or gravitation.

\section{Isometric embeddings into $\R^{3,1}$}
Consider an isometric embedding of $\sigma$ as $X:S^2\rightarrow \R^{3,1}$ with components $X=(X_0, X_1, X_2, X_3)$
such that \[ -(dX_0)^2+\sum_{i=1}^3 (dX_i)^2=\sigma.\] Denoting the image surface $X(\Sigma)$ by $\Sigma_0$, we also consider the projection $\widehat{\Sigma}$ given by $(X_1, X_2, X_3): S^2 \rightarrow \R^3$. The induced metric on $\widehat{\Sigma}$  is $\sigma+(dX_0)^2$. The above observation motivates the following procedure of constructing isometric embeddings.  Take any function $\tau$ on $(S^2, \sigma)$, $\sigma+(d\tau)^2$ gives another Riemannian metric. 
As long as  the Gauss curvature of $\sigma+(d\tau)^2$ is positive, one can first isometrically embed 
$\sigma+(d\tau)^2   $ into $\R^3$ with image $\widehat{\Sigma}$, and then take the graph of $\tau$ over $\widehat{\Sigma}$ in $\R^{3,1}$.
The space of isometric embedding into $\R^{3,1}$ is essentially parametrized by such time function $\tau$.  But which $\tau$ is the best? 

\section{The definition of quasilocal mass with $\R^{3,1}$ reference}

 It turns out the surface Hamiltonian satisfies the following gravitational conservation law that can be used to relate surface Hamiltonians on $\Sigma_0$ and $\widehat{\Sigma}$.  \begin{equation}\label{conservation} \int_{\widehat{\Sigma}} {\widehat{H}} = \int_{\Sigma_0}[ { |H_0|} {\sqrt{1+|\nabla\tau|^2}} \cosh\theta_0-\nabla \theta_0\cdot\nabla \tau -{\alpha_{H_0}}(\nabla\tau)],\end{equation} where $\theta_0$ is given by
\[\sinh\theta_0=\frac{-\Delta\tau}{\sqrt{1+|\nabla\tau|^2}{ |H_0|}}.\] Here $H_0$ is the mean curvature vector field of $\Sigma_0$ in $\R^{3,1}$, and $\alpha_{H_0}$ is the connection one-form determined by $H_0$.

 The identity \eqref{conservation} provides a way to motivate the quasilocal energy definition of Wang-Yau \cite{Wang-Yau1, Wang-Yau2}.  Give a spacelike 2-surface $\Sigma$ in a physical spacetime with the data $(\sigma, |H|, \alpha_H)$. Let $X$ be an isometric embedding of $\sigma$ into $\R^{3,1}$ with time function $\tau$. Let $\widehat{\Sigma}$ be the 
projection of $\Sigma_0=X(\Sigma)$, the quasilocal energy is defined to be:
\[8\pi E(\Sigma, X, \tau)=\int_{\widehat{\Sigma}} \widehat{H}- \int_{\Sigma}[ { |H|} {\sqrt{1+|\nabla\tau|^2}} \cosh\theta-\nabla \theta\cdot\nabla \tau -{ \alpha_{H}}(\nabla\tau)] \] where $\theta$ is given by
\begin{equation}\label{sinh}\sinh\theta=\frac{-\Delta\tau}{\sqrt{1+|\nabla\tau|^2}{ |H|}}.\end{equation}
 For a surface in the Minkowski spacetime, the above conservation law \eqref{conservation} implies the corresponding quasilocal energy is zero. 
 After the positivity \cite{Wang-Yau1, Wang-Yau2} is established, one can minimize among the time functions to define the quasilocal mass. The Euler-Lagrange equation is called the 
optimal isometric embedding equation:

\begin{equation}\label{**}
\begin{split}&({\widehat{H}\hat{\sigma}^{ab} -\hat{\sigma}^{ac} \hat{\sigma}^{bd} \hat{h}_{cd}})\frac{\nabla_b\nabla_a \tau}{\sqrt{1+|\nabla\tau|^2}}\\
&= div_\sigma (\frac{\nabla\tau}{\sqrt{1+|\nabla\tau|^2}} \cosh\theta{ |H|}-\nabla\theta-{ \alpha_{H}})\end{split}\end{equation}
with $\sinh\theta$ given by \eqref{sinh}.

Coupling the last equation with the isometric embedding equation into $\R^{3,1}$, we obtain four equations for the four unknown coordinate
functions $(\tau, X_1, X_2, X_3)$. The solution gives the optimal isometric embedding and produces a surface in $\R^{3,1}$ that is supposed to be the ``best match" for the physical surface \cite{Chen-Wang-Yau2}. We can then define other quasilocal conserved quantities such as quasilocal angular momentum based on this reference surface \cite{Chen-Wang-Yau3}.

% The second variation of the quasilocal energy was also studied by Miao-Tam-Xie and Chen-Wang-Yau and local minimizing and rigidity properties were proved for certain critical points. For example, when $\alpha_H=0$, $\tau=0$ or an isometric embedding into $\R^3$ is a ``stable" critical point, i.e. $\delta^2 E\geq 0$.  This actually leads to a new inequality for surfaces in $\R^3$. 

%
%

\section{The definition of quasilocal mass with static reference}
 There have been recent progresses  \cite{Chen-Wang-Yau4, Chen-Wang-Wang-Yau1} to replace the reference Minkowski spacetime by a static reference spacetime. A static spacetime is a time-oriented Lorentzian 4-manifold (with possibility nonempty smooth boundary) such that there exists a static chart 
$(t, x^1, x^2, x^3)$  under which the metric takes the form
\begin{equation}\label{metric_form}
\check{g}= -V^2(x^1, x^2, x^3) dt^2 +\bar{g}_{ij} (x^1, x^2, x^3) dx^i dx^j,
\end{equation}
where the static potential $V$ satisfies $V>0$ on the interior and $V=0$ on the boundary.

 A static slice ($t=\text{constant}$) is a Riemannian 3-manifold with the induced metric $\bar{g}=\bar{g}_{ij}$. For simplicity, we assume the spacetime is vacuum with possibly nonzero cosmological constant and thus $V$ satisfies the static equation
\[\bar{\Delta} V \bar{g} - \bar{\nabla}^2 V + V \bar{Ric} =0.\]  Isometric embeddings of 2-surfaces into our examples such as the de Sitter, Anti-de Sitter, and Schwarzschild spacetimes have been studied by  Chang-Xiao \cite{Chang-Xiao}, Lin-Wang \cite{Lin-Wang}, Guan-Lu \cite{Guan-Lu}, etc.

 Again for a 2-surface $\Sigma$ in a physical spacetime with geometric data $(\sigma, H)$, we consider an isometric embedding $X$ of $\sigma$ into a reference static spacetime time with time function $\tau$ (with respect to $T_0=\frac{\partial}{\partial t}$) and the projection $\hat{\Sigma}$ on a static slice (with mean curvature  $\hat{H}$), the quasilocal energy $E(\Sigma, X, T_0)$ \cite{Chen-Wang-Yau4, Chen-Wang-Wang-Yau1} is defined to be 

\[\begin{split}
   & \frac{1}{8 \pi}  \int_{\hat{\Sigma}} V \widehat H - \frac{1}{8 \pi}
 \int_\Sigma  \Big [ \sqrt{(1+ V ^2| \nabla \tau|^2) |H|^2   V ^2 + div( V ^2 \nabla \tau)^2 }  \\
& \qquad -   div( V ^2 \nabla \tau)  \sinh^{-1} \frac{ div( V ^2 \nabla \tau) }{ V  |H|\sqrt{1+ V ^2| \nabla \tau|^2} }
  -  V ^2 \alpha_{H} (\nabla \tau)  \Big ].
\end{split}
\] We note that when $V=1$ this is exactly the expression for the quasilocal energy with respect to Minkowski reference. For a surface in the static spacetime, the expression vanishes by a gravitational conservation law similar to \eqref{conservation}.

% Introducing  $\rho$ and $j_a$ in this case: 
%\[ \begin{split}\rho &= \frac{\sqrt{|H_0|^2 +\frac{(div V^2 \nabla \tau)^2}{V^2+V^4 |\nabla \tau|^2}} - \sqrt{|H|^2 +\frac{(div V^2 \nabla \tau)^2}{V^2+V^4 |\nabla \tau|^2}} }{ V\sqrt{1+ V^2|\nabla \tau|^2}}. \end{split}\]
%\[ j_a=V^2 \rho \nabla_a \tau-\nabla_a \sinh^{-1}\frac{\rho div (V^2\nabla\tau)}{|H_0| |H|}-(\alpha_{H_0})_a+(\alpha_H)_a\]

% $E(\Sigma, X, T_0)$ can be rewritten as
%\[E(\Sigma, X, T_0)=\frac{1}{8\pi} \int_\Sigma
%V^2 ( \rho+j_a \nabla^a \tau) .\]

% For any Killing field $K$ on the reference static spacetime, we define the corresponding quasilocal conserved quantity: \[E(\Sigma, X, T_0, K)=-\frac{1}{8\pi} \int_\Sigma
%\left[ \langle K, T_0\rangle \rho+j(K^\top) \right],\]
%where $K^\top$ is the tangential part of $K$ to $X(\Sigma)$,

%
%
% The first variation of qusailocal energy with respect to a static spacetime with static potential $V$ is
%The is rewritten from the equation in Theorem 2.8 of arXiv:1604.02983
%\[ \delta E=\frac{1}{8\pi}\int_\Sigma \delta (V^2) [\frac{1}{2} \rho -(\nabla^a\tau) j_a ]-\frac{1}{8\pi} \int_\Sigma \delta \tau \nabla^a (V^2 j_a).\] 

%Note that $\delta (V^2) $ and $\delta \tau$ are constraint by the linearized isometric embedding equation. 

 One can check that for any spacelike surface in the reference static spacetime, the variation of $E$ vanishes. When the reference isometric embedding $X$ lies in a static slice (i.e. $\tau=0$), the quasilocal mass expression becomes
\[ \frac{1}{8\pi} \int V(H_0-|H|) d\Sigma.\] The Brown-York-Liu-Yau mass corresponds to the special case with $\R^3$ reference when $V=1$. Combing with the quasi-spherical construction of Bartnik-Shi-Tam \cite{Bartnik1, Shi-Tam}, the positivity of $\int V(H_0-H)$ with respect to the $\mathbb{H}^3$ reference is proved by Wang-Yau \cite{Wang-Yau3} and Shi-Tam \cite{Shi-Tam2}. Recently, Lu-Miao \cite{Lu-Miao} established a positivity theorem with respect to the Schwarzschild reference that leads to the proof of a quasilocal Penrose's inequality. In retrospect, such a consideration is natural as  the model spacetime for Penrose's inequality in which the equality holds is exactly the Schwarzschild spacetime.
\section{Rigidity properties}
 I would like to discuss the rigidity property in the rest of this article. The following theorem corresponds to the simplest version, which can be interpreted as a spacetime infinitesimally rigidity theorem of isometric embedding.
% The kind of positivity and rigidity property we can prove in the static reference case are much weaker than the Minkowski reference case.

 \begin{thm}\cite{Chen-Wang-Yau2}
Let $(\sigma, |H|, \alpha_H)$ be the data of a spacelike surface $\Sigma$ in the Minkowski spacetime. Suppose the projection of $\Sigma$ onto the orthogonal complement of $T_0$ is convex. Then the kernel of the linearized optimal isometric 
embedding system consists precisely of Lorentz motions. Moreover, the second variation of the quasilocal energy $E(\Sigma, \tau)$
is non-negative definite. \end{thm}

 Comparing with Cohn-Vossen's rigidity theorem of isometric embedding of convex surfaces in $\mathbb{R}^3$, the theorem implies that surfaces
with prescribed $(\sigma, |H|, \alpha_H)$ are infinitesimally rigid in $\mathbb{R}^{3,1}$. The identification of the kernel of the linearized optimal embedding system enables the solution of the optimal isometric embedding system in a perturbative configuration by an inverse function argument.

The above theorem remains true for a convex surface in a static slice of de Sitter or Anti-de Sitter spacetime, see the next theorem.  However, a new phenomenon occurs for the Schwarzschild spacetime.

\begin{dfn}
An isometric embedding into a static slice is infinitesimally rigid if the kernel of the linearized isometric embedding equation consists of the restriction of the Killing vector fields of the static slice to the surface.
\end{dfn}

\begin{rem}
Every convex surface in $\R^3$, $\SSS^3$, or $\HH^3$ is infinitesimally rigid. However, this is no longer true for convex surfaces in a static slice of the Schwarzschild spacetime. In fact, even a  round sphere of symmetry admits isometric deformations that are not induced by any ambient isometry, see Li-Wang \cite{Li-Wang}. On the other hand,
there are rigidity theorems for isometric embeddings into a static slice of the Schwarzschild spacetime by Chen-Zhang \cite{Chen-Zhang} and Li-Miao-Wang \cite{Li-Miao-Wang} under an additional condition.
\end{rem} Isometric embedding theorems into the Schwarzschild spacetime were established by the work of Lu \cite{Lu1}, Guan-Lu \cite{Guan-Lu}, and Li-Wang \cite{Li-Wang}.

Since convex surfaces in $\SSS^3$ and $\HH^3$ are infinitesimally rigid, we have the following spacetime rigidity theorem for the de Sitter and Anti-de Sitter spacetimes. 
\begin{thm} \cite{Chen-Wang-Wang-Yau1} The spacetime infinitesimal rigidity theorem holds true for convex surfaces in a static slice of the de Sitter or Anti-de Sitter spacetime.\end{thm}

Namely $\delta^2 E\geq 0$ and the kernel
of the linearized optimal isometric embedding consists of the global isometry of the ambient spacetime. 
Such a rigidity theorem should be enough to anchor a unique solution of the optimal isometric embedding system for a perturbative configuration of the background spacetime.

\noindent {\bf Acknowledgement:} This material is based upon work supported by the National Science Foundation under Grants No. DMS-1405152 and No. DMS-1810856 (Mu-Tao Wang).

%%%%%%%%%%%%%%%%%%%%%%%%%%%%%%%%%
% References
%%%%%%%%%%%%%%%%%%%%%%%%%%%%%%%%%

\end{document}